\documentclass[acus]{JAC2003}

\usepackage{graphicx}

\usepackage{tabularx}
\usepackage{multirow}
\usepackage{units}
\usepackage[position=bottom,skip=2pt]{caption}
\usepackage{comment}
\usepackage{booktabs}
\usepackage{caption}
\usepackage{amsmath}
\usepackage{textcomp}

\newcommand{\Gross}[2]{
	\begin{figure*}[t]
		\begin{center}
			\includegraphics{TUPSO24_fig#1.pdf}
			\caption{#2}
		\label{#1}
	\end{center}
\end{figure*}}

\newcommand{\Klein}[2]{
	\begin{figure}[h]
		\begin{center}
			\includegraphics{TUPSO24_fig#1.pdf}
			\caption{#2}
		\label{#1}
	\end{center}
\end{figure}}

\begin{document}

\title{DISPERSION BASED BEAM TILT CORRECTION}
\author{Marc W. Guetg\thanks{marc.guetg@psi.ch}, Bolko Beutner, Eduard Prat, Sven Reiche\\ Paul Scherrer Institut, CH-5232 Villigen PSI, Switzerland}
\maketitle

\section{Abstract}
In Free Electron Lasers (FEL), a transverse centroid misalignment of longitudinal slices in an electron bunch reduces the effective overlap between radiation field and electron bunch and therefore the FEL performance. The dominant sources of slice misalignments for FELs are the incoherent and coherent synchrotron radiation within bunch compressors as well as transverse wake fields in the accelerating cavities. This is of particular importance for over-compression which is required for one of the key operation modes for the SwissFEL planned at the Paul Scherrer Institute.

The centroid shift is corrected using corrector magnets in dispersive sections, e.g. the bunch compressors. First and second order corrections are achieved by pairs of sextupole and quadrupole magnets in the horizontal plane while skew quadrupoles correct to first order in the vertical plane.
Simulations and measurements at the SwissFEL Injector Test Facility are done to investigate the proposed correction scheme for SwissFEL. This paper presents the methods and results obtained.

\section{Introduction}
FELs need high-current low-emittance ($\varepsilon$) beams to lase. The slice emittance $\varepsilon_\text{slice}$ is of special importance. But the projected emittance $\varepsilon_\text{proj}$ influences the gain as well, since the effective overlap between electrons and radiation field is decreased.

Measurements of the slice parameter are more difficult since streaking is required. If projected and slice parameters are similar the operation of any accelerator is simplified.

\subsection{SwissFEL}
The SwissFEL injector will use an S-band gun followed by 6 S-band cavities (with a laser heater after the first two) and 2 X-band cavities followed by a bunch compressor BC ($R_{56} = -55.1$~mm) with 330~MeV nominal beam energy for operation \cite{freq,Brown,CDRSwi}. The injector is followed by a C-Band linac bringing the beam up to 2.1~GeV and the second BC ($R_{56} \geq -22.5$ mm) compressing up to 3~kA. The final C-Band linac boosts the energy up to 5.8 GeV and is followed by an energy collimator with variable $R_{56}$ (nominally at zero). 

With the exception of the laser heater all magnetic chicanes are symmetric and horizontal, and all of them contain a pair of quadrupole, skew quadrupole and sextupole magnets each to correct for the centroid shifts. 

For longitudinal diagnostic there is a transverse deflection cavity (TDC) (50~cm up to 4.9 MV) after the first BC as well as the BC quadrupoles and skew quadrupoles to streak the beam. 

A schematic overview in Fig.~\ref{1} shows the setup. Since SwissFEL is not yet built all results concerning SwissFEL are obtained by simulation using elegant \cite{elegant}.

\Klein{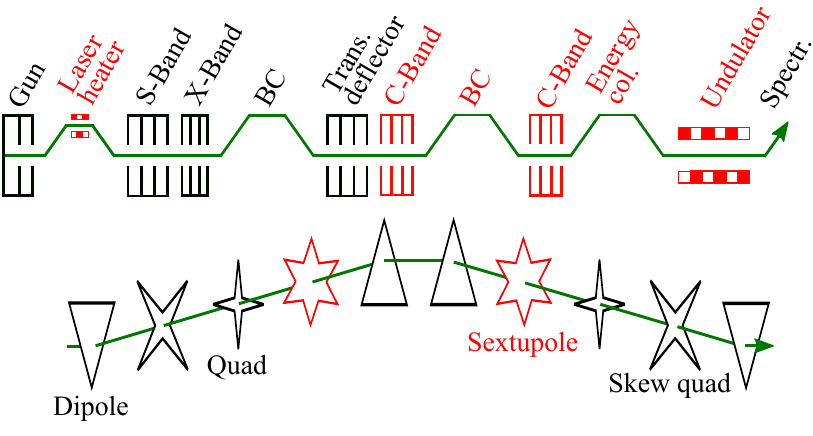}{Schematic drawing of SwissFEL. SITF consists only of the black elements. The lower part is a more detailed sketch of the bunch compressors. The green arrow corresponds to the electron beam.}
 
\subsection{SwissFEL Injector Test Facility}
The SwissFEL Injector Test Facility (SITF) implements the injector of the future SwissFEL facility. It is built for testing components and beam parameters for SwissFEL. 

The SITF consists of an S-band gun, four S-band and one X-band cavities for acceleration up to 270~MeV on-crest. We reduced the energy to 180~MeV due to off-crest acceleration and linearisation of the longitudinal phase space. SITF has the same longitudinal diagnostic options as described for SwissFEL.

There are no sextupoles installed at the SITF. 
%A schematic overview in Fig.~\ref{1} shows the setup whereby only the black elements are there. 
The black elements in Fig.~\ref{1} show the SITF setup.
The SITF beam line setup was used for measurements as well as simulations \cite{CDRInj}.

\section{Beam Dynamics}
We use the statistical definition of the emittance $\varepsilon_\text{rms}$ for simulations and for measurements Gaussian fits are used to obtain the beam sizes. It has been shown that both methods are equivalent in our case (mostly Gaussian beams).
The variable $\varepsilon$ denotes projected and normalized emittance. 

\Gross{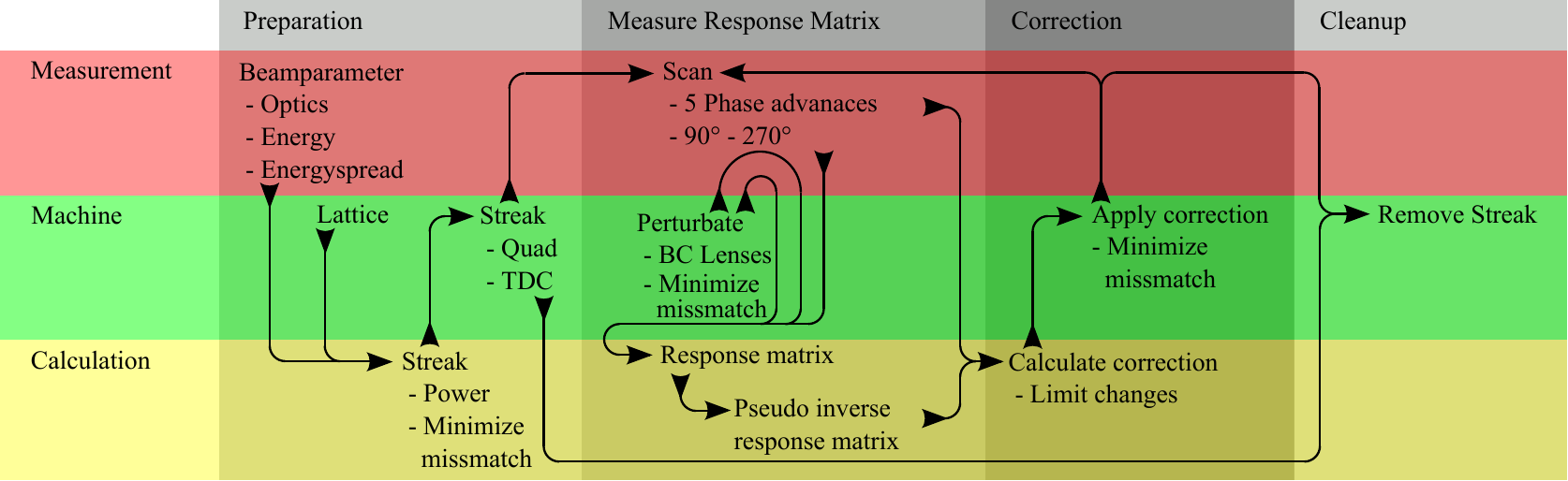}{Simplified flowchart of the correction algorithm. See text for detailed information.}

\subsection{The Parameter $\chi$}
To characterize the centroid shift we introduce the multi-order complex-valued parameter $\chi$ for each transverse plane. 
We use $x$ for the transverse axis and $z$ for the longitudinal axis. To be independent of $\varepsilon$ and optics functions beam coordinates are normalized with respect to their standard deviations. Transverse centroid positions $x_c$ and $x'_c$ along $z$ propagate according to the Courant-Snyder theory. This allows a definition of $\chi$ similar to $\varepsilon$, where the magnitude stays constant in non-dispersive adiabatic systems.

Re($\chi$) represents any introduced $x'/z$ kick which is translated through betatronic phase-advance into Im($\chi$), the spatial displacement of slices along the bunch. Arg($\chi$) corresponds to the phase advance between source and observation point.

The Taylor expansion of $\chi$ linking spatial and momentum centroid displacements along the bunch is given by
\begin{equation}
	\frac{x'_c(z)}{\sigma_{x'}} + \frac{x_c(z)}{\sigma_x}\cdot i = \sum_{n=0}^\infty \chi_n \left(\frac{z}{\sigma_z}\right)^n
	\label{chi}
\end{equation}

\subsection{Manipulation of $\chi$}
Changing $\chi$ in a well defined manner requires a non-zero dispersion $\eta$ and a linear longitudinal phase space. Both are available within the bunch compressors. To set a specific $\chi_{1,\text{hor}}$ two quadrupoles are used to leak out a well defined $\eta$ according to
\begin{equation}
	\Delta(\eta_\text{hor}'\delta) + \Delta x_{\text{hor},\beta}' = \frac{\eta_\text{hor}^0\delta + x_{\text{hor},\beta}^0 }{f_\text{quad}}
	\label{Q}
\end{equation}
where $f$ corresponds to the focal length of the quadrupole and $\delta$ to the energy spread. The $\beta$ subscript indicates the betatronic position. Since a quadrupole in a dispersive section only introduces Re($\chi$) two quadrupoles are needed, separated by a phase advance which is not a multiple of $\pi$. The manipulation of $\chi_{1,\text{ver}}$ is done analogously using two skew quadrupoles:
\begin{equation}
	\Delta(\eta_\text{ver}'\delta) + \Delta x_{\text{ver},\beta}' = \frac{\eta_\text{hor}^0\delta + x_{\text{hor},\beta}^0}{2f_\text{skew}}
	\label{SQ}
\end{equation}
For SwissFEL there are also two pairs of sextupoles available to correct $\chi_{2,\text{hor}}$ according to
\begin{multline}
	\Delta(\eta_\text{2,hor}'\delta^2) + \Delta\xi_\text{hor}' + {\cal O}((x_{\text{hor},\beta}^0)^2 + (x_{\text{ver},\beta}^0)^2) =\\ m\left[(x_{\text{ver},\beta}^0)^2 - (\eta_\text{hor}^0\delta + x_{\text{hor},\beta}^0)^2\right] 
	\label{S}
\end{multline}
where $m$ is the focal strength of the sextupole and $\eta_2$ corresponds to the second-order dispersion. The resulting chromaticity $\xi$ is used to correct for natural chromaticity. The leaked first-order vertical dispersion is corrected for by the skew quadrupoles.

\section{Algorithm}

\subsection{Streaking}
To measure the transverse centroid shift in one plane the longitudinal phase space needs to be mapped onto the orthogonal transverse plane. This is achieved by two different methods.

For the simplest case we use the transverse deflection cavity TDC to map the beam longitudinal coordinate $z$ onto $y$. This will not be possible after the second bunch compressor of SwissFEL since there is no TDC foreseen in that position.

An alternative way is to use one of the (skew) quadrupole magnets in the BC to leak out dispersion $\eta$ and thus mapping energy onto the respective transverse plane. For linear longitudinal phase space dispersion-based streaking proved to be viable since it works in both transverse planes.

\subsection{Measurement}
For the SITF $\chi$ is measured using 5 equidistant betatron phase-advance steps in the range $\left[\frac{\pi}{2},\frac{3\pi}{2}\right]$ in the correction plane and constant phase advance in the streaking plane ($\frac{\pi}{2}$). The $\beta$ functions in both transverse planes are kept constant at the location of the screen. Ten bunches were recorded to reduce the impact of shot-to-shot fluctuations.

\subsection{Correction}
A schematic flow chart of the correction algorithm is given in Fig.~\ref{2}. The beam parameters need to be known to measure $\chi$ and correct for the mismatch in the optical functions. For mild mismatches in energy and optics the correction algorithm still works, but correction of the mismatch and accurate measurement of the initial $\chi$ is not possible.

After the initial beam setup the electron bunch is streaked by one of discussed methods (TDC / quadrupole / skew quadrupole) and the resulting mismatch (not for TDC) is then corrected. This preparatory step has to be done only once. 

The response matrix is then measured perturbating the quadrupoles / skew quadrupoles / sextupoles of the measurement plane recording the penalties $\chi_1$, $\chi_2$ and chromaticity $\xi$. For SITF no sextupoles were used and no $\xi$ was measured. Quadrupoles and skew quadrupoles for SwissFEL are used to minimize the introduced mismatch. For SITF only quadrupoles are used since there are no skew quadrupoles installed in the high energy section.

For the SITF also a direct calculation of the needed $\eta$ and $\eta'$ with subsequent application to the beam was performed successfully. This method offers the benefit of being much faster and better scalable to more knobs but has the disadvantage of being more sensitive to errors in optics / energy / orbit in the bunch compressor.

The pseudo-inverse of the measured response matrix is then used for several iterations of correction. This is viable since the introduced changes are small. In case of instabilities and machine drifts remeasuring the response matrix may be needed, however. To avoid instabilities in the correction, any changes in the magnets are limited in magnitude to not more than twice the value of the perturbation.

When finished the system has to be optimized by removing the streak and rematching the optics.

\subsection{Simulation Results for SwissFEL}

To test the performance of the correction the three main operation modes of SwissFEL were considered. These three operation modes consist of Short Pulse mode (SP) for ultra short FEL pulses, Long Pulse mode (LP) being the default mode and Large Bandwidth mode (LB) using over-compression to increase electron energy spread leading to large photon bandwidth. All modes are further described in Table~\ref{OpMode}. To simulate the offsets in the cavities they were displaced randomly by a Gaussian distribution ($\sigma_\text{off}$ = 500~\textmu m) in both transverse planes.

\begin{table}[h!tbp]\centering \renewcommand{\arraystretch}{1.2}
	%\begin{tabular}{lrrrrrr}
	\caption{The tree main operation modes for SwissFEL to benchmark the correction algorithm. $\Delta\varepsilon$ is significantly reduced for all the cases. CF stands for compression factor.}
	\begin{tabular}{p{1.8cm} | p{0.6cm} p{0.6cm} | p{0.6cm} p{0.6cm} | p{0.6cm} p{0.6cm}}
		\hline
		\hline
		%\toprule
		 & \multicolumn{2}{c|}{SP} &  \multicolumn{2}{c|}{LP} &  \multicolumn{2}{c}{LB}\\
		\cline{2-7}
		& w/o & w & w/o & w & w/o & w\\
	    \hline
		$|\chi_x|$ [$10^{-3}$] & 670 & 53 & 826 & 13 & 557 & 109\\
		$|^2\chi_x|$ [$10^{-6}$] & \textless 0.1 & \textless 0.1 & 448 & 3& 3048 & 779\\
    	$|\chi_y|$ [$10^{-3}$] & 387 & 3 & 167 & 0.65 & 422 & 7\\
    	$|^2\chi_y|$ [$10^{-6}$] & \textless 0.1 & \textless 0.1 & 15 & 1 & 311 & 20\\
		$|\xi_x|$ & 0.63 & 0.59 & 0.05 & 0.12 & 0.59 & 0.68\\
		$\Delta\varepsilon_x$ [\%]& 39 & 6.2 & 111 & 13 & 349 & 83\\
     	$\Delta\varepsilon_y$ [\%]& 19 & 4.6 & 11 & 9.8 & 18 & 9\\
     	$\sigma_z$ [\textmu m] & 1.1 & 1.3 & 5.5 & 5.4 & 5.1 & 8.3\\	
		%\hline
		$\varepsilon_x$ [nm]& \multicolumn{2}{c|}{104} & \multicolumn{2}{c|}{354} & \multicolumn{2}{c}{354} \\
		$\varepsilon_y$ [nm]& \multicolumn{2}{c|}{104} & \multicolumn{2}{c|}{354} & \multicolumn{2}{c}{354} \\
		Charge [pC]	& \multicolumn{2}{c|}{10} & \multicolumn{2}{c|}{200} & \multicolumn{2}{c}{200} \\
		CF & \multicolumn{2}{c|}{310} & \multicolumn{2}{c|}{153} & \multicolumn{2}{c}{--164} \\
		% \bottomrule
		\hline
		\hline
	\end{tabular} 
	\label{OpMode}
\end{table}

The results in Table~\ref{OpMode} clearly show the effectiveness of the tilt correction for the SwissFEL configurations. The growth in projected emittance could be reduced significantly for all the modes. 

The biggest $|\chi|$ was observed for the LB mode which has the highest charge density within the second BC due to over-compression. The transverse offset of the cavities was set to larger misalignment values than nominal tolerances to stress-test the stability of the algorithm. The specific case of the LB mode is presented in Fig.~\ref{3}.

\Klein{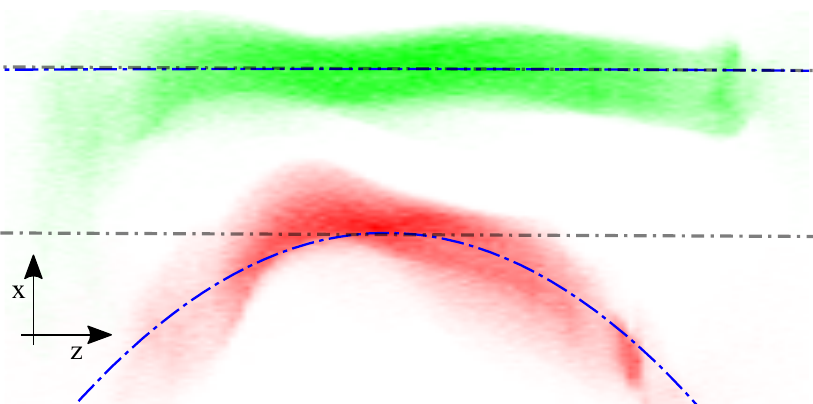}{Uncorrected (red, bottom) and corrected (green, top) beam for the LB mode. (The displacement in $x$ is introduced to separate the distributions for better visibility.)}

\subsection{Measurements at the SITF}

To benchmark the correction algorithm the SITF was used with various compression factors and different offsets in the X-band cavity. Due to the  machine status at the time of the measurement the charge was limited to 85~pC for the wake field measurements and to 140~pC for the compression study.

\Klein{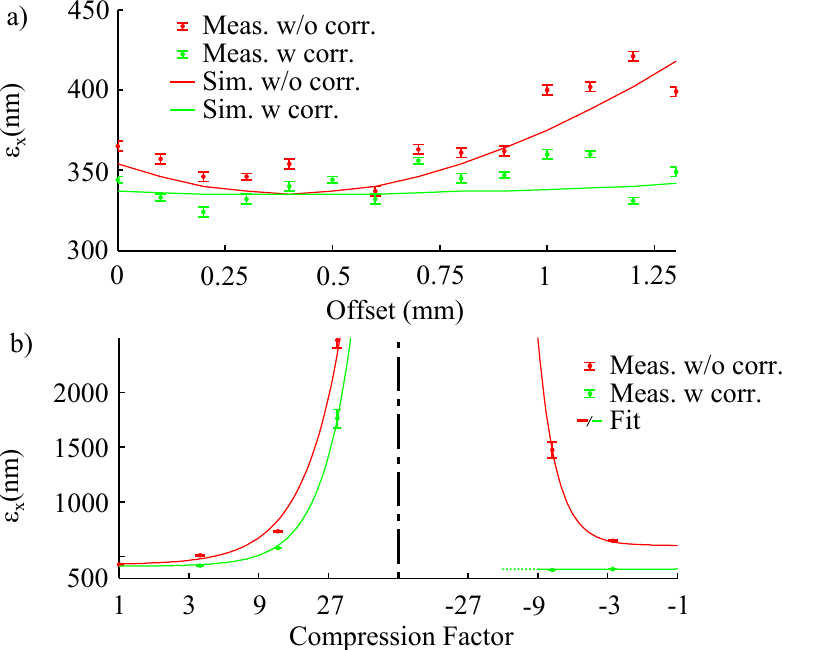}{Results of the measurements for horizontal wake field measurement and compression study (a and respectively b).}

Figure~\ref{4} shows a significant reduction of the projected emittance independent of the source of the tilt. One specific phase-advance measurement using over-compression is shown in Fig.~\ref{5}.

\Klein{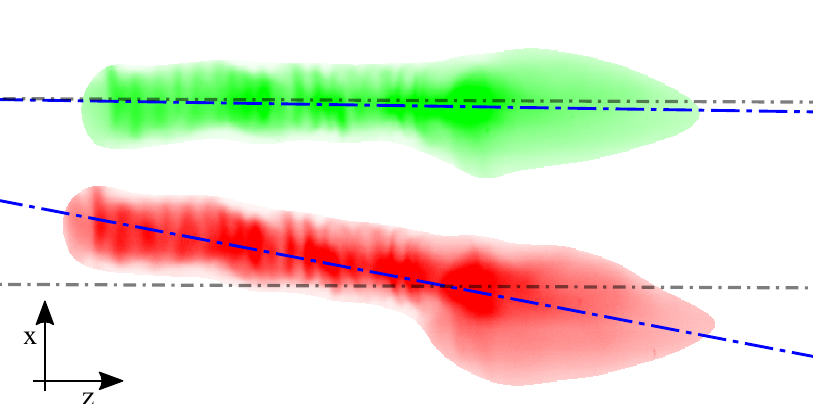}{Streaked beams obtained by measurement whereby bottom red beam corresponds to the uncorrected beam and the top green one to the correct situation. (The displacement in $x$ is introduced to separate the distributions for better visibility.)}

\section{Discussion}
Our method is very robust in terms of optical, energy and positioning (of BC magnets) mismatch. The correction of $\chi$ is a relatively cheap method to increase FEL gain compared to other approaches or extension of the undulator line. It therefore plays an important role in any FEL design.

Correcting $\chi$ in addition to $\xi$ has the benefit of improving the operability of the accelerator since projected parameters are much easier to measure than slice parameters.

It is crucial to have a unified parameter to quantify the centroid shift when comparing different systems. $\chi$ proved to be an appropriate choice.

\subsection{Issues}

The proposed correction algorithm is time consuming (about 10--20 min for both planes at the SITF). This effect can be mitigated by reusing the response matrix, thereby effectively reducing the time in half. The time needed can be even further decreased by creating operation tables for given settings used in a single iteration.

The skew quadrupoles introduce minor coupling which could only be corrected for by solenoids near the gun at the SITF. For SwissFEL there are skew quadrupoles in non-dispersive sections to minimize coupling.

The optical elements in the BC introduce an energy-dependent orbit jitter. Monte Carlo simulations have shown that with the given amplitude/phase tolerances of the RF system the resulting orbit jitter is within our specification of less than a tenth of the beam size.

The change in $R_{56}$ given by 
\begin{equation}
	R_{56} = \int\limits_{BC} \frac{\eta(s)}{\rho(s)}ds ,
	\label{R56}
\end{equation}
where $\rho$ denotes the bending radius of the dipoles,
and $T_{566}$ within the BC changes the compression factor \cite{Edwards}.
This can be accounted for by either changing the angle of the BC or the RF settings to keep compression constant. The shape of the current profile was mostly unchanged by the applied correction.
\bibliographystyle{unsrt}

\begin{thebibliography}{6}
	\bibitem{freq} H.~Meinke, ``Taschenbuch der Hochfrequenztechnik'', 4.~Auflage, Springer-Verlag, 1985.
	\bibitem{Brown} K.~Brown, ``A First- and Second-Order Matrix Theory for the Design of Beam Transport Systems and Charged Particles Spectrometers'', SLAC-75, 4, 1982.
	\bibitem{CDRSwi} R.~Ganter, ``SwissFEL Conceptual Design Report'', 2012.
	\bibitem{elegant} M.~Borland, ``elegant: A Flexible SDDS-Compliant Code for Accelerator Simulation'', Advanced Photon Source LS-287, 2000.
	\bibitem{CDRInj} M.~Pedrozzi, ``SwissFEL Injector Conceptual Design Report'', 2010.
	\bibitem{Edwards} D.~Edwards, ``An Introduction to the Physics of High Energy Accelerators'', Wiley-vch, 2004.
	%\bibitem{pseudo} Penrose, A generalized inverse for matrices, Proceedings of the Cambridge Philosophical Society,51, 1955.
	%\bibitem{LBWmode} S.~Reiche, Status of the SwissFEL Facility at the Paul Scherrer Institute, proc.~FEL'12, Shanghai, 2012.
\end{thebibliography}

%% Bib %%

\end{document}